\documentclass[twocolumn,prl,superscriptaddress]{revtex4}
\usepackage{graphics}
\usepackage{graphicx}
\usepackage{amsfonts}
\usepackage{textcomp}
\usepackage{amssymb}
\usepackage{mathrsfs}
\usepackage{amsmath}
\usepackage{color}
\usepackage{float}

\begin{document}

\title{Comment on ``Explicit Analytical Solution for Random Close Packing in $d=2$ and $d=3$''}

\author{Duyu Chen}
\email[correspondence sent to: ]{duyu@alumni.princeton.edu}
\affiliation{Materials Research Laboratory, University of California, Santa Barbara, California 93106, United States}
\author{Ran Ni}
\email[correspondence sent to: ]{r.ni@ntu.edu.sg}
\affiliation{School of Chemical and Biomedical Engineering, Nanyang Technological University, Singapore 637459} 


\maketitle
In Ref. \cite{Za22}, the author presented an explicit analytical derivation of the volume fractions $\phi_{\rm RCP}$ for random close packings (RCP) in both $d=2$ and $d=3$. Here we first briefly show the key parts of the derivation in Ref.~\cite{Za22}, and then provide arguments on why we think the derivation of the analytical results is problematic and unjustified.

In this comment, we use the $d=3$ case as an example to briefly show the derivation for $\phi_{\rm RCP}$ following Ref.~\cite{Za22} and the issues associated with the derivation, but we note that similar issues apply to $d=2$ as well. It has been well known \cite{To18} that pair correlation function $g(r)$ for a jammed hard-sphere packing can be decomposed into a discrete probability distribution $g_{\rm c}(r)$ from particle contacts and a continuous-function contribution in the region beyond contacts $g_{\rm bc}(r)$:
\begin{equation}
\label{eq1} g(r) = g_{\rm c}(r)+g_{\rm bc}(r).
\end{equation}
Specifically, $g_{\rm c}(r)$ is given by
\begin{equation}\label{eq2}
g_{\rm c} (r) = g_0 g(\sigma) \delta (r - \sigma),
\end{equation}
where $g(\sigma)$ is the contact value of $g(r)$, and $g_0$ is a normalization factor to be determined later. These two equations are Eqs. 5 and 6 in Ref.~\cite{Za22}, respectively.

In jammed states, the coordination number or the average number of particles in contact with a test particle can be calculated as
\begin{equation}\label{coord}
z = 4 \pi \rho \int_{0}^{\sigma^+} g(r) r^2 dr,
\end{equation}
where $\rho \equiv 6 \phi /(\pi \sigma^3) $ is the density of the jammed state, and $\sigma^+ \equiv \sigma + \epsilon$ with $\epsilon \rightarrow 0^+$. This is the Eq. 3 in Ref.~\cite{Za22}.  The author in Ref.~\cite{Za22} used a known jammed structure, i.e., the closest packing face-centred-cubic (FCC) crystal of $\phi_{\rm CP} = \pi/(3\sqrt{2})$ and $z_{\rm CP} = 12$, as a reference state combined with Eqs.~\ref{eq1},~\ref{eq2}, and \ref{coord} and the Percus-Yevick (PY) equation of state (for calculating $g_{\rm CP}(\sigma)$) \cite{So89, Le84}, to obtain the normalization factor $g_0 \approx 0.0331894 \sigma$. With these, the author used $z_{\rm RCP} = 6$ to obtain $\phi_{\rm RCP} \approx 0.658963$ in Eq. 12 in Ref.~\cite{Za22}. Another similar value of $\phi_{\rm RCP} \approx 0.677376,$ was also obtained in Ref.~\cite{Za22} using the Carnahan-Starling (CS) equation of state \cite{So89}. 

As one can see here, the accuracy of the analytical estimate of $\phi_{\rm RCP}$ relies heavily on the accuracy of the equation of state used to compute $g_{\rm CP/RCP}(\sigma)$ and $g_0$. 
We note that the PY and CS equation of state were derived primarily for \emph{equilibrium hard-sphere fluids}~\cite{Fr01}, and using them for jammed crystals, e.g., FCC crystal, like in Ref.~\cite{Za22}, has never been justified and may cause huge errors. For example, if one uses the calculated $g_0$ in Ref.~\cite{Za22} combined with the PY equation of state to compute the packing fraction of another jammed crystal, i.e., the close packed simple cubic (SC) crystal with $z_{\rm SC} = 6$, one would arrive at an estimate of $\phi_{\rm SC} \approx 0.658963$, which is the same as $\phi_{\rm RCP}$ in Ref.~\cite{Za22}, while it is known that the correct value should be $\phi_{\rm SC} = \pi / 6 \approx 0.5236$. This clearly shows that the way of estimating $g_{\rm CP}(\sigma)$ and $g_0$ is problematic, which is simply because that the PY or CS equation of state has only one input $\phi$ and it does not distinguish between different structures at the same $\phi$. Therefore, we believe that the estimate of $\phi_{\rm RCP}$ in Ref.~\cite{Za22} is also problematic, as it is based on an incorrectly estimated $g_0$ and $g_{\rm CP}(r)$; the fact that the theoretical estimate falls within the large range of empirical values from past computer simulations is largely fortuitous; and the author’s claim of solving $\phi_{\rm RCP}$ analytically in Ref.~\cite{Za22} is highly misleading. 

Lastly, we would like to mention another unjustified way of using the PY and CS expressions in Ref.~\cite{Za22}. 
The pair correlation function from PY or CS expressions $g^{\rm PY/CS}(r)$ does not have any Dirac-$\delta$-function contribution, and is finite at contact $r=\sigma$ for $\phi<1$; consequently, if we plug $g^{\rm PY/CS}(r)$ into $g(r)$ in Eq.~\ref{coord}, we would always get $z = 0$ for any jammed state, which is unphysical and violates the central assumption that the derivation in Ref.~\cite{Za22} relies on. Ultimately, this is because the PY or CS expressions cannot be directly used in jammed states. On the other hand, the author in Ref.~\cite{Za22} used the PY and CS expressions intended for the estimation of the total $g(r)$ at $r=\sigma$ to estimate $g(\sigma)$ in $g_{\rm c}(r)$ (Eq.~\ref{eq2}), 
which we believe is an inappropriate use of the PY or CS expressions. To the best of our knowledge, this way of using $g^{\rm PY/CS}(\sigma)$ in Eq.~\ref{eq2} (Eq. 6  in Ref.~\cite{Za22}) has never been done before, and we fail to see any physical justification.

\section{Addendum} 
Zaccone recently published an Erratum ~\cite{Za22Err} to mention that the analytical solution to RCP in Ref.~\citep{Za22} is \emph{not exact} but rather an approximation. Here we would like to emphasize that the Erratum does not address or clarify the concerns raised in this Comment. In the Erratum, it says \emph{``possible inaccuracies in the approximation for $g(\sigma^+)$ are indeed effectively compensated by the choice of an ordered reference state to determine the unknown dimensionful constant $g_0$''}, and the author mentioned four different values of $g_0$ based on PY+fcc, PY+bcc, CS+fcc, and CS+bcc yielding an RCP from 0.643 to 0.677 in 3D, which is close to the numerical value around 0.64  previously obtained from computer simulations. There are two issues here. First, as we mentioned above, PY/CS are equation of states for equilibrium hard-sphere fluid, and they have never been and should not be used for hard-sphere crystals. Because it is well known that fluid and crystals are completely different phases, and their equations of state are qualitatively different from each other, which can been seen from the literature on the equation of state for hard spheres, e.g.,~\cite{halleos}. Therefore, to use PY/CS to approximate $g(\sigma^+)$ of crystals, e.g., fcc/bcc, for calculating $g_0$ is 
 \emph{not inaccurate but fundamentally wrong}. Second, in the Erratum~\cite{Za22Err}, Zaccone showed the obtained RCP values using fcc/bcc as the ordered reference state, which immediately triggers an important question: \emph{why choose fcc/bcc rather than other close packing crystals, e.g., simple cubic, as the ordered reference state?} One can find that if choosing simple cubic as the ordered reference state, using the method in Ref.~\citep{Za22}, we will obtain $\phi_{\rm RCP} = \pi/6 \approx 0.5236$, which is very far from the acceptable range of RCP in the scientific community. Does this suggest that the purpose of choosing fcc or bcc as the ordered reference state in Refs.~\cite{Za22,Za22Err} is merely to match the numerical value of RCP around 0.64 in three dimensions? 
 

\end{document}